# Dielectric study on mixtures of ionic liquids

E. Thoms[1†], P. Sippel[1]*, D. Reuter[1], M. Weiß[1], A. Loidl[1,2] & S. Krohns[1,2]

[1]Experimental Physics V, Center for Electronic Correlations and Magnetism, University of Augsburg, 86135 Augsburg, Germany
[2]Institute for Materials Resource Management, University of Augsburg, 86135 Augsburg, Germany

*Correspondence and requests for materials should be addressed to P.S. (pit.sippel@physik.uni-augsburg.de)

**Ionic liquids are promising candidates for electrolytes in energy-storage systems. We demonstrate that mixing two ionic liquids allows to precisely tune their physical properties, like the dc conductivity. Moreover, these mixtures enable the gradual modification of the fragility parameter, which is believed to be a measure of the complexity of the energy landscape in supercooled liquids. The physical origin of this index is still under debate; therefore, mixing ionic liquids can provide further insights. From the chemical point of view, tuning ionic liquids via mixing is an easy and thus an economic way. For this study, we performed detailed investigations by broadband dielectric spectroscopy and differential scanning calorimetry on two mixing series of ionic liquids. One series combines an imidazole based with a pyridine based ionic liquid and the other two different anions in an imidazole based ionic liquid. The analysis of the glass-transition temperatures and the thorough evaluations of the measured dielectric permittivity and conductivity spectra reveal that the dynamics in mixtures of ionic liquids are well defined by the fractions of their parent compounds.**

Ionic liquids became one of the most popular material classes in material science[1,2,3,4] based on their versatile potential applications[5,6,7,8,9]. They are salts that are liquid below 100 °C and offer beneficial properties, e.g., low volatility and high electrochemical stability. This makes them ideal candidates for solvent-free electrolytes to improve many energy applications[8]. Unfortunately, ionic liquids do not satisfy the requirements for energy-storage technologies in terms of their ionic conductivity[10], which is a key-factor in supercapacitors and batteries. However, they are regarded as 'designer solvents' because of the vast number of possible anions and cations to form an ionic liquid that is extended by the possibility to combine two or three ionic liquids via mixing[11]. There is a high possibility to design an ionic-liquid electrolyte that meets the required ionic conductivity[6,10,12]. Tuning ionic liquids via mixing is desirable, as it is more economic than altering their molecular structure[13]. Interestingly, an increased conductivity has been reported for some mixtures of ionic liquids (e.g., refs. 14,15,16) in comparison to their parent compounds. These and similar findings suggest that unique properties are accessible via mixing ionic liquids[17]. This way of optimization seems to be restricted to only a few substances. For many mixing series the concentration is linear to the conductivity in a logarithmic scale[4]. Most of the series analyzed so far closely follow ideal mixing-laws[13,18,19], although the general picture is still unclear[4]. Niedermeyer *et al.* and Chatel *et al.* (refs. 4 and 17, respectively) review the topic in detail and point out the necessity for more data sets of the physical and chemical properties (e.g., conductivity) for ionic liquid mixtures. A systematic use of spectroscopic analysis is required to understand the complex intrinsic interactions and the underlying physical mechanisms[17], which are of key relevance for future applications based on mixtures of ionic liquids[17]. To the best of our knowledge, there are only a few works considering the temperature dependence of the conductivity of ionic liquid mixtures (i.e. refs. 14,16,20,21,22,23,24) and solely Stoppa *et al.*[15] performed dielectric spectroscopy. In the present work, we thoroughly investigate two binary mixtures of ionic liquids using broadband dielectric spectroscopy and differential scanning calorimetry (DSC). The mixtures are analyzed regarding their glass-transition temperature ($T_g$), ionic conductivity and their main reorientational relaxation, as determined from the dielectric spectra via an equivalent-circuit approach. Finally, we verify the correlation of $T_g$ and the fragility index $m$ to the conductivity, which was found in a previous work[25]. In the first series, a two cation-based ionic liquid of 1-Butyl-3-methyl-imidazolium tetrafluoroborate (BMIM BF$_4$) and 1-Butylpyridinium tetrafluoroborat (BPY BF$_4$) is examined. For the second series a two anion-based ionic liquid of 1-Butyl-3-methyl-imidazolium bis(trifluoromethylsulfonyl)imide (BMIM TFSI) and 1-Butyl-3-methyl-imidazolium chloride (BMIM Cl) is used. Since water impurity may influence the physical properties of ionic liquids[26,27], all samples were dried prior to the measurement and the amount of water was determined via Karl Fischer titration (KFT) when possible.

---

[†] Present address: Division for Biophysics and Molecular Physics, Silesian Center for Education and Interdisciplinary Research, 75 Pulku Piechoty 1A, 41–500 Chorzow, Poland

## Results and Discussion

**Differential Scanning Calorimetry.** Upon cooling, most ionic liquids can be solidified via a glass transition[25,28] as it was observed for all samples by DSC. Fig. 1 (a) and (b) show a magnified view of the step-like anomaly in the heat flow that is associated with the glass transition from the heating run for $BPY_xBMIM_{1-x}$ $BF_4$ and BMIM $Cl_xTFSI_{1-x}$, respectively. The measurements were performed between 300 K and 100 K with a cooling and heating rate of 10 K/min. Prior to this measurements, all samples were dried under suitable conditions. The preparation of the samples, drying procedures and water contents are described in detail in the methods section. To determine the glass-transition temperature, the onset method is used, as depicted by the blue dotted straight lines in Fig. 1 (a) and (b).

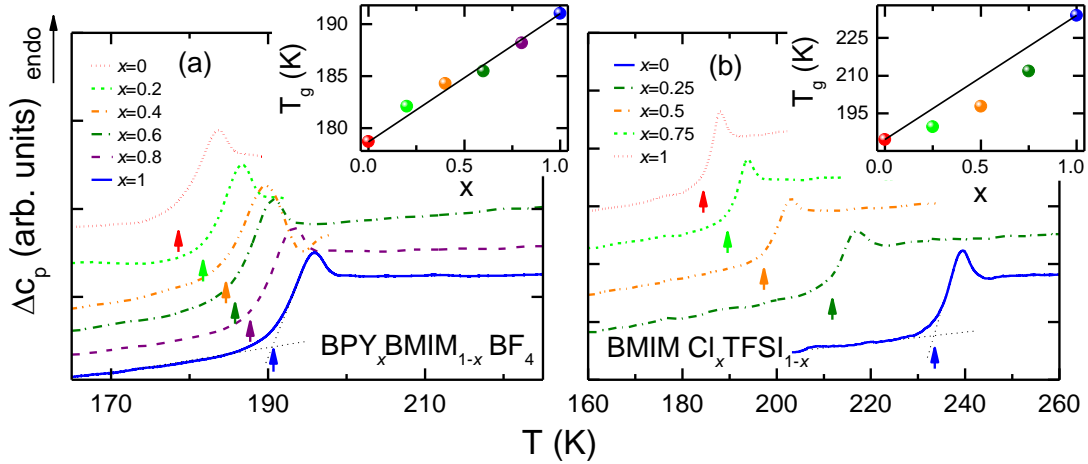

**Figure 1. Shift of glass-transition temperature for two mixtures of ionic liquids.** Differential Scanning Calorimetry heating traces of $BPY_xBMIM_{1-x}$ $BF_4$ (a) with $x$ = 0/0.2/0.4/0.6/0.8/1.0 and BMIM $Cl_xTFSI_{1-x}$ (b) with $x$ = 0/0.25/0.5/0.75/1.0 (endotherm up). The colored arrows mark the onset points used to determine $T_g$. The dependence of $T_g$ on composition is shown in the insets, where the black lines indicate linear extrapolation of $T_g$ between $x$ = 0 and 1.

The glass-transition temperature for the pure component BPY $BF_4$ of the series of $BPY_xBMIM_{1-x}$ $BF_4$ (Fig.1 (a)) is 191 K. When adding BMIM $BF_4$, $T_g$ shifts continuously to lower temperatures as indicated by the arrows. Finally, for pure BMIM $BF_4$, $T_g$ is 179 K. The inset in Fig. 1 (a) illustrates the change of $T_g$ as function of the mole fraction $x$ of $BPY_xBMIM_{1-x}$ $BF_4$. This change can be described by the line that represents the weighted average of $T_g$ of the two pure ionic liquids, as the deviations are well below 1 K for all samples. Likewise, a glass-transition temperature between the two simple ionic liquids of a mixture has been reported in literature[21,29] and many ionic-liquid mixtures reveal a nearly linear trend of $T_g$ with a change of concentration[13,14].

The second mixing series of BMIM $Cl_xTFSI_{1-x}$, is shown in Fig. 1 (b). BMIM Cl exhibits a $T_g$ of 234 K and again, $T_g$ is lowered with decreasing $x$. For pure BMIM TFSI we report a $T_g$ of 185 K. The black line in the inset of Fig. 1 (b) assumes a linear change in the glass-transition temperature. However, the measured transition temperatures of the samples with $x$ = 0.25, 0.5 and 0.75 show a decrease of 7, 11 and 9 K with respect to a linear extrapolation between the glass-transition temperatures of the two constituents. We showed previously[27] that residual water has a huge impact on $T_g$ of BMIM Cl. In the present case, the glass-transition temperature of BMIM Cl has a strong impact on the ideal-mixing line. We assume that water impurities cannot explain this discrepancy. Overall, the glass-transition temperatures are between the parent compounds. Since all glass-transition temperatures of the mixtures are below the linear extrapolation, it is possible that the huge TFSI-anion dominates the glass transition.

**Dielectric Spectroscopy.** We have measured the dielectric response of the two mixing series $BPY_xBMIM_{1-x}$ $BF_4$ with x = 0/0.6/1.0 and BMIM $Cl_xTFSI_{1-x}$ with $x$ = 0/0.5/1.0 in a broad frequency-range from 1 Hz to 3 GHz (for BPyBF4 the spectrum was expanded to 40 GHz) and a wide temperature range from 170 to 373 K. This range allows a thorough analysis of the conductivity and the reorientational modes in the glassy and liquid phases of these systems. The samples were dried at about 373 K for at least 16 h inside a nitrogen gas cryostat until the dielectric properties reach time-dependent constant values (not shown). This is a typical sign for no further water removal[27], suggesting stable conditions with low values of water impurities. Subsequently, the samples were measured in the same cryostat to avoid a reabsorption of water prior to the measurement. The extended measurements above 3 GHz (c.f. Fig. 2 (g-i)) had to be performed in ambient atmosphere and therefore previous drying in an oven in analogy to the calorimetry measurements were performed. Unfortunately, in all cases the sample size is too small for KFT, thus the exact water content cannot be determined after the dielectric measurements. However, no differences between the temperature-dependent cooling and heating runs were found, suggesting a constant and low amount of water impurity. Fig. 2 displays the spectra of the cooling run of the $BPY_xBMIM_{1-x}$ $BF_4$ series for selected temperatures. The real ((a), (d) and (g)) and imaginary ((b), (e) and (h)) parts of the dielectric permittivity ($\varepsilon'$ and $\varepsilon''$, respectively) as well as the conductivity $\sigma'$ in (c), (f) and (i) are shown. The latter emphasizes the dc conductivity, although the information is already contained in $\varepsilon''$ because of $\sigma' = \omega\varepsilon''$. The rather controversially debated modulus representation (i.e., $M^* \propto 1/\varepsilon^*$), sometimes used to characterize the ionic dynamics in ion conductors[30,31,32,33,34] is not presented, since permittivity and conductivity allow a more direct determination of the technically relevant dc conductivity.

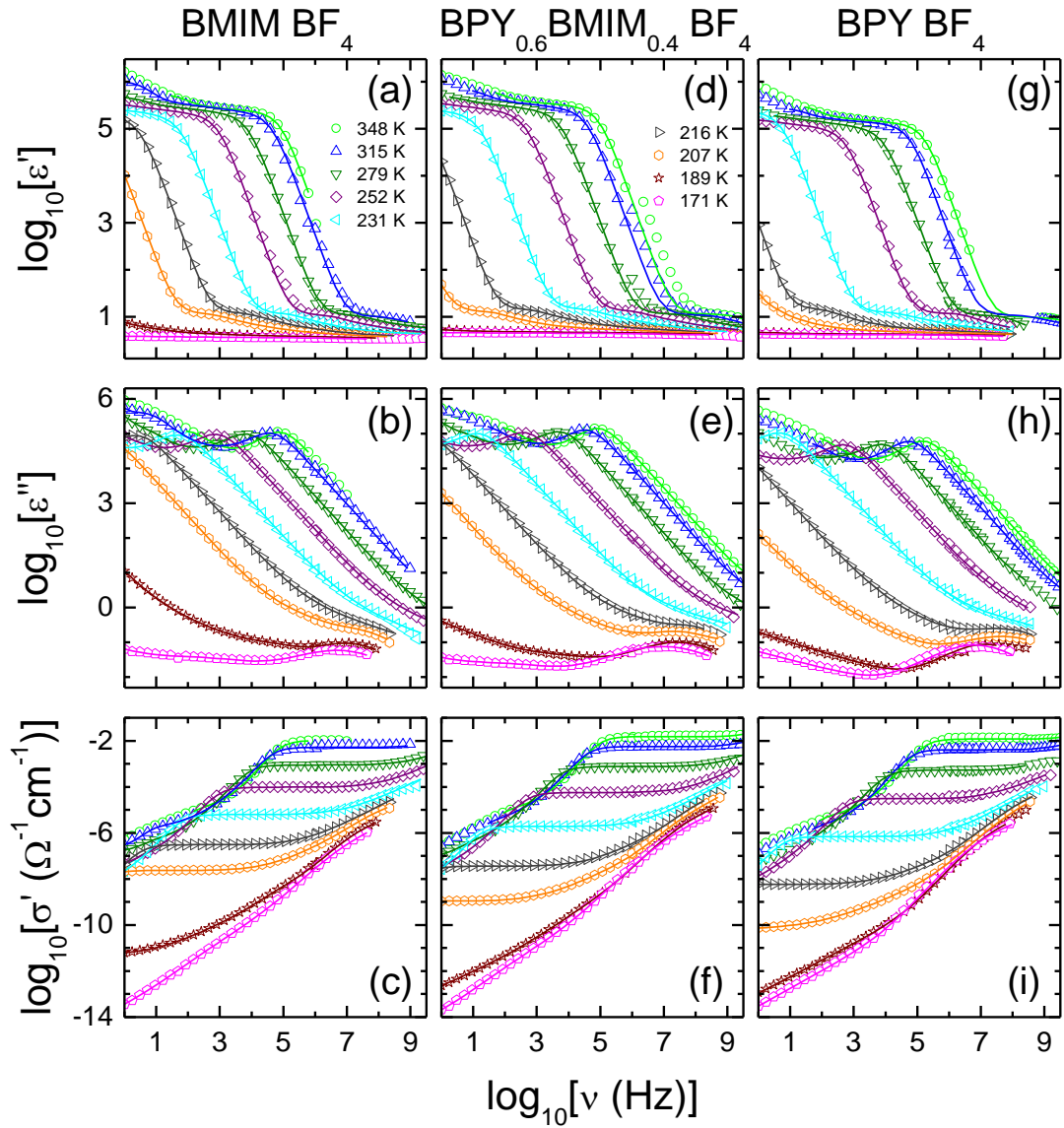

**Figure 2. Dielectric spectra of ionic liquids and their mixture.** Frequency dependence of permittivity $\varepsilon'$, dielectric loss $\varepsilon''$ and conductivity $\sigma'$ of BMIM $BF_4$ (a)-(c), $BPY_{0.6}BMIM_{0.4}$ $BF_4$ (d)-(f) and BPY $BF_4$ (g)-(i). The lines are simultaneous fits of $\varepsilon'$ and $\varepsilon''$ with two distributed RC circuits[35] in series with dc conductivity and the sum of three intrinsic relaxations[36].

The spectra of Fig. 2 (a) are dominated by a huge increase of $\varepsilon'(\nu)$ above $10^5$ towards lower frequencies, which is typically found in many ion conductors, including ionic liquids[35,37]. This non-intrinsic effect of electrode polarization leads to so-called colossal values[38] in $\varepsilon'$ that superimposes the intrinsic sample properties. The electrode polarization shows up as a so-called Maxwell-Wagner (MW) relaxation step in $\varepsilon'(\nu)$ that should lead to well-defined plateaus at low frequencies[36]. However, as in many ion conducting materials, no distinct plateau is approached[35]. This effect can be ascribed by the presence of several MW-relaxations or by a distribution of MW-relaxation times as discussed, e.g., in ref. [35]. The MW-relaxation evolves as a step in $\varepsilon'$ and a local maximum in $\varepsilon''$, is observed in Fig. 2 (b), e.g., at about $10^5$ Hz for the 348 K curve and is shifting towards lower frequencies with decreasing temperature. This temperature dependence originates from an increasing sample viscosity. The high-frequency flanks of these local maxima arise from the dc conductivity, which becomes more obvious in the conductivity representation (Fig. 2 (c)) by the corresponding frequency-independent plateaus, e.g., from 1 kHz to 1 MHz for the 231 K curve. At lower frequencies, the electrode polarization leads to a decrease in conductivity, making it necessary to conduct ac-measurements to determine the dc conductivity.

At high frequencies, e.g., above 1 MHz for the 231 K curve, the conductivity in Fig. 2 (c) has a pronounced frequency dependency. This indicates the presence of one or more intrinsic relaxations. The dielectric loss (Fig. 2 (b)) shows signatures of a relaxational process indicated as peaks at the lowest temperatures and at elevated frequencies, e.g., around 5 MHz for the 171 K curve. The corresponding steps in the real part of the dielectric permittivity (Fig. 2 (a)) are not visible at the current scaling of the spectra. Fig. 2 (a) reveals additional shoulders that are superimposed on the electrode polarization effect, indicating a further intrinsic relaxation with $\varepsilon_s$ in the order of 10, e.g., above 1 kHz for the 252 K curve. The corresponding peaks in the dielectric loss (Fig. 2 (b)) are covered by the dc conductivity but their high-frequency flank may cause the flattening of the spectra towards higher frequencies. This is also indicated by the increase of the conductivity at frequencies above the dc-plateau (Fig. 2 (c)).

The processes described above can be analyzed in more detail by fitting the spectra simultaneously in $\varepsilon'(\nu)$ and $\varepsilon''(\nu)$ with an equivalent-circuit[36,38]. To account for the strong MW-relaxations, we used two distributed RC circuits[35,39], connected in series to the bulk response. The latter consists of a resistor representing the dc conductivity, connected in parallel to three complex capacitances that represent the intrinsic reorientational modes. In addition to the two relaxations mentioned above, at some temperatures a third process, allocated at frequencies in-between, improved the accuracy of the fit significantly. Although the whole equivalent-circuit represents the sample, only parts of the elements were needed for the fit at certain temperatures. The main reorientational mode, e.g., in Fig. 2 (a) at 1 kHz for the 207 K curve, was modeled by a Cole-Davidson (CD) function[40], as often employed in glassy matter[41,42]. It is most likely caused by the reorientation of the cations[43,44]. The other relaxations were modeled by Cole-Cole functions[45]. For imidazole-based ionic liquids, a relaxation similar to the relaxational process at higher frequencies, e.g., around 5 MHz for the 171 K curve in Fig. 2 (b), was discussed[43] in terms of a secondary Johari-Goldstein relaxation[46]. The Cole-Cole function is known to provide a good description for such secondary relaxations[47]. In summary, the found spectra are typical for ionic liquids and similar succession of dynamic processes were reported in several works[25,43,44,48,49,50,51,52,53,54].

In general, the dielectric properties of BPY$_{0.6}$BMIM$_{0.4}$ BF$_4$ and BPY BF$_4$, shown in Fig. 2 (d)-(f) and (g)-(i), are quite similar to the spectra of BMIM BF$_4$. Again, the equivalent-circuit described above was used to analyze the dielectric behavior. For BPY BF$_4$, the third intrinsic process was not required to describe the experimental data. The origin of this third intrinsic relaxation is still under debate. Our results are suggesting that it is only present in imidazole-based ionic liquids. However, the relaxation is rather weak ($\Delta\varepsilon \approx 0.2$) and superimposed by stronger relaxations for the most temperatures. Thus, we cannot exclude that the third relaxation is ubiquitous for all investigated ILs.

The dielectric properties of BMIM Cl$_x$TFSI$_{1-x}$ (not shown) can be well described with the same equivalent-circuit. The third intrinsic relaxation was also found in all liquids of this series and it is most pronounced for BMIM TFSI leading to $\Delta\varepsilon \approx 0.9$. For BMIM TFSI, this relaxation was attributed in Ref. 43 to an intrinsic Johari-Goldstein process arising from the presence of a nonsymmetrical anion. However, for ILs with symmetric anions, e.g. BF$_4$, this intrinsic relaxation emerges, too. A conventional Johari-Goldstein relaxation, as discussed for BPY$_{1-x}$BMIM$_x$ BF$_4$, can explain the dielectric behavior.

For the BPY$_{1-x}$BMIM$_x$ BF$_4$ series, the dc conductivity is changing with respect to $x$ at lower temperatures. While at temperatures above 279 K (cf. Fig. 2 (c), (f) and (i)) $\sigma_{dc}$ reveals almost the same value when changing $x$, at lower temperatures the plateau significantly diverges, e.g., at 207 K in Fig. 2 (c), (f) and (i). This implies a change in the temperature dependence of the dc conductivity based on different glass-transition temperatures of these liquids. Naturally the local maximum in $\varepsilon''$ (cf. Fig. 2 (b), (e) and (h)) of the MW-relaxation shifts towards lower frequencies for reduced conductivities. At lower temperatures (T < 235 K) the main reorientational mode (cf. Fig. 2 (a) around 1 kHz for 207 K) is affected as well, which is shown in Fig. 2 (d) and (g). All properties of the investigated mixture are located well between their two parent ionic liquids, which is also expected from the measurements of the glass-transition temperatures. The discussion focuses on the temperature dependence of the dc conductivity and the main reorientational mode determined from the fits of the dielectric properties for both mixing series.

**Relaxation and conductivity dynamics.** Fig. 3 shows the average relaxation time $\langle\tau_\alpha\rangle = \beta\tau_\alpha$ (where $\beta$ is the broadening parameter of the CD function) of the main relaxation ((a) and (c)) and the dc conductivity ((b) and (d)) for both measurement series in an Arrhenius representation, as obtained by the fits described above. Both, the main relaxation time and the dc conductivity reveal the typical non-Arrhenius temperature dependence known from glass forming systems. Such behavior was reported for many ionic liquids[25,32,34,50,53,55,56,57,58,59,60,61]. Fits (solid lines) with the empirical Vogel-Fulcher-Tammann (VFT) law[62,63,64,65],

$$\tau = \tau_0 \exp\left[\frac{-DT_{VFT}}{T-T_{VFT}}\right], \quad (1)$$

where $\tau_0$ is the inverse attempt frequency, $T_{VFT}$ the divergence temperature and $D$ the so-called strength parameter[65], are most suitable to describe the obtained data. For the dc conductivity, a modified form of eq. (1) with a negative exponent and $\sigma_0$ as prefactor instead of $\tau_0$ was used. Most commonly, the fragility index $m$, which is defined as the slope at $T_g$ in the Angell plot, $\log \tau$ vs. $T_g/T$ (ref. 66), is used to parameterize the deviation from Arrhenius behavior. Following Böhmer et al. (Ref. 66), it can be calculated from the strength parameter via $m = 16 + 590/D$. The VFT-fits of $\langle\tau_\alpha\rangle$ allow to determine $T_g$ using the empirical relation $\langle\tau_\alpha\rangle(T_g) = 100$ s. In many ionic liquids the glass-transition temperature of the ionic subsystem $\tau_\sigma(T_g) = 100$ s, derived from the modulus relaxation, reasonably matches the temperature of $\sigma_{dc} = 10^{-15}$ $\Omega^{-1}$ cm$^{-1}$ (ref. 25). Here, we determined $T_g$ of the ionic subsystem from the conductivity via this relation. These two temperatures are directly evaluated by the intersection of the VFT-fit and the upper ordinates (c.f. Fig. 3). Generally, the main relaxation time and the dc conductivity of both parent compounds comprise the behavior of their mixtures over the whole temperature range, as illustrated by Fig. 3.

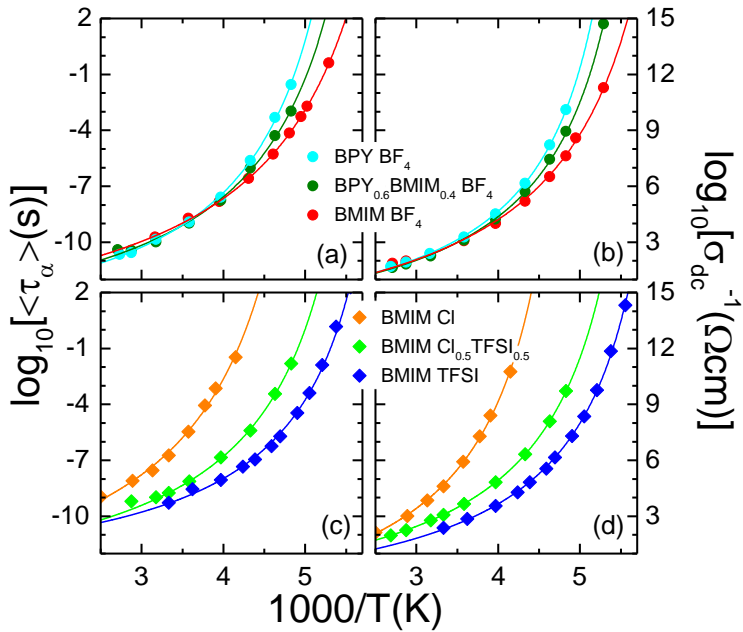

**Figure 3. Temperature dependent ionic dynamics of BPY$_x$BMIM$_{1-x}$ BF$_4$ and BMIM Cl$_x$TFSI$_{1-x}$.** The relaxation times (a) and (c) and the dc-conductivities (b) and (d) are shown in an Arrhenius representation. The data are determined from the fits of the dielectric spectra. The lines are fits with the Vogel-Fulcher-Tammann-equation[62,63,64,65].

For BPY$_x$BMIM$_{1-x}$ BF$_4$ (Fig. 3 (a) and (b)), the evolution of the dc conductivity and the relaxation times reveal almost identical values above 286 K (1000/T < 3.5). Due to their different glass-transition temperatures, the curves diverge upon further cooling. The glass-transition temperatures of the main relaxational mode for BPY BF$_4$, BPY$_{0.6}$BMIM$_{0.4}$ BF$_4$ and BMIM BF$_4$ are 197, 191 and 182 K, respectively. The temperatures deduced from the conductivity are slightly (about 2 K) below. The transition temperature of the mixture is at the weighted average of $T_g$ of the two pure parent ionic liquids, which was also evidenced by the DSC measurements. However, the absolute values are slightly above the DSC results. This indicates a different water content for both experiments. Another explanation is a decoupling effect of the structural glass-transition temperature derived from DSC measurements and the ionic or rotational subsystems. We expect, based on our previous work (ref. 25), a change in fragility due to the similar dc-conductivities at room-temperature and different glass-transition temperatures. Indeed, the fragility index $m$, determined from the conductivity, is decreasing from 117 via 111 to 93 for lower glass-transition temperatures. This mixing series is close to ideal mixing as described by Clough et al. in Ref. 13 for their ionic liquid mixtures.

The dielectric properties of the pure ionic liquids of BMIM $Cl_x TFSI_{1-x}$ differ even at high temperatures, as shown in Fig. 3 (c) and (d). BMIM Cl exhibits the highest $T_g$. Its dynamics are slower than those of BMIM TFSI and the mixture. The glass-transition temperatures determined from the main relaxational mode are 181, 195 and 232 K for BMIM TFSI, BMIM $Cl_{0.5}TFSI_{0.5}$ and BMIM Cl, respectively. The glass-transition temperatures deduced from the conductivity are 179, 191 and 228 K, which is slightly lower. Corresponding to the results of the DSC measurement, the mixture exhibits a $T_g$ which is about 10 K lower than the expected $T_g$ using the weighted average of $T_g$ of the pure ionic liquids. In this regard, this ionic liquid mixture deviates from the ideal mixing behavior. However, the fragility index $m$ of BMIM $Cl_{0.5}TFSI_{0.5}$, determined from the VFT-fit of the dc conductivity, is about 107. Interestingly, this is close to the weighted average of fragility indices of the parent compounds BMIM TFSI and BMIM Cl, being 120 and 97, respectively.

**Correlation of $\sigma_{dc}$ with m and glass-transition temperature.** Many aprotic ionic liquids exhibit a correlation of room-temperature conductivity with their $T_g$ and fragility index[25]. This correlation is deduced from the temperature dependent dc conductivity following a VFT-law, which is dominated by these two glass-parameters. The room temperature resistivity is calculated by inserting the relaxation time of the glass-transition temperature ($\tau(T_g) = 100$ s) and the formula for the fragility index ($m = 16 + 590/D$) into the VFT-equation[25]. The resistivity derived from that formula is shown as color-coded plane in the background of Fig. 4. This illustrates the trend of the room-temperature resistivity in correlation with $T_g$ and $m$. As indicated by the red arrow the resistivity increases with increasing $T_g$ and decreasing fragility. $T_g$ and the fragility of both mixing series were determined from the VFT-fit of the dc conductivity, as discussed above.

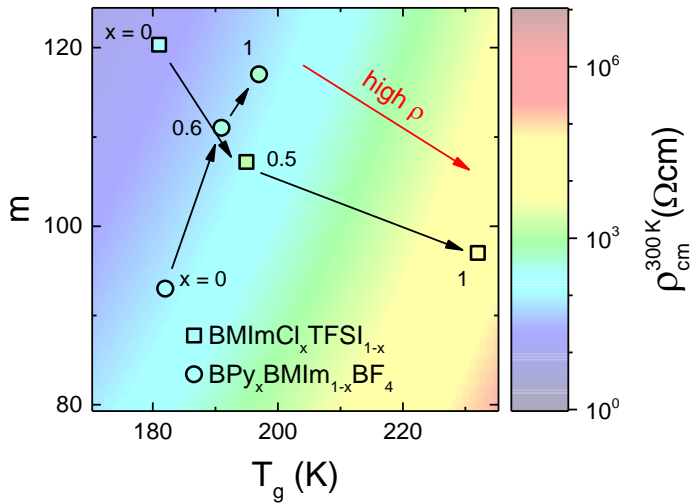

**Figure 4. Fragility, glass-transition temperature and room-temperature resistivity of the mixed ionic liquids.** Fragility and $T_g$ for $BPY_x BMIM_{1-x} BF_4$ and BMIM $Cl_x TFSI_{1-x}$, derived from the temperature dependence of the ionic dynamics. Black arrows indicate variations of $x$. The color-coded background represents resistivity as a correlation of $m$ and $T_g$, according to ref. 25. The color symbols represent the measured value of the resistivity of the samples.

The circles in Fig. 4 represent the results of $BPY_x BMIM_{1-x} BF_4$. The color of the symbols, which depicts the measured resistivity, shows the almost identical dc conductivity at room-temperature. The connecting black arrows that indicate variations of $x$, are almost parallel to a line of constant resistivity given by the $\rho(T_g,m)$ relation, following a trajectory of isoresistivity. That means, changes of the resistivity from the reduced $T_g$ are compensated by the fragility, as it is expected from the Arrhenius plot in Fig. 3. The squares represent BMIM $Cl_x TFSI_{1-x}$, which shows an increasing resistivity for increasing $x$. Both the decreased fragility index and the increased $T_g$ enhance the resistivity. The variation of $x$ results in resistivities almost parallel to the gradient of the colored plane. In summary, the trend of the dc-resistivities is well described by the colored plane, allowing to characterize the conductivity of the ionic liquids mixtures via the fragility index and $T_g$.

**Summary**
Ideal mixing of the ionic liquids BPY $BF_4$ and BMIM $BF_4$ is shown. We provide glass-transition temperatures revealed by DSC experiments and by dielectric spectroscopy of $BPY_{0.6}BMIM_{0.4} BF_4$, corresponding to the weighted average of their parent compounds. In addition, the dielectric properties of the mixture mimic those of the parent compounds, exhibiting relaxational dynamics and conductivities exactly between them. Interestingly, even the fragility index of the mixture, which can be regarded as a measure for the complexity of the energy landscape, is well between the parent compounds. The dielectric properties of BMIM $Cl_x TFSI_{1-x}$ suggest a linear behavior of the fragility when mixing the two parent compounds. This indicates the possibility to design ternary ionic liquids with precisely defined fragilities, opening a way to gain further insights on a quantity whose physical origin is still unclear. However, the linearity of the fragility when mixing ionic liquids has to be evidenced in further compounds. In contrast, the glass-transition temperatures of the mixtures deviate about 10 K from their expected value, which is obtained by the weighted average of $T_g$ of the pure ionic liquids. It is still unclear if the strong molecular difference of the anions is the reason. The relaxational dynamics and dc-conductivities of all investigated mixtures studied in this work are between the parent ionic liquids. This enables a simple route to tune , e.g., the conductivity as well as other physical properties of ionic liquids. Particularly, this way to modify the properties of ionic liquids is not only very precise, but also very economic compared to a functional design of new anion cation pairs[13]. Our work on two ionic liquid mixture series gains further insights, especially it relates the dielectric properties to the concentration.

**Methods**
**Sample preparation.** The ILs were purchased from IoLiTec with a purity of 99%. The mixtures were prepared by blending the two neat ILs in the appropriate mass ratio. To minimize water content, all samples were dried in $N_2$-gas or vacuum at elevated temperatures for at least 16 hours prior to the measurements. The water content of $BPY_xBMIM_{1-x} BF_4$ was determined using coulometric KFT after drying but before to the DSC measurements. The mixing series exhibits less than 0.3 mol% residual water for all samples: $BMIM\ BF_4$ 0.23 mol%, $BPY_{0.6}BMIM_{0.4}\ BF_4$ 0.25 mol% and $BPY\ BF_4$ 0.27 mol%. The high viscosity and hygroscopic nature of the samples of the $BMIM\ Cl_xTFSI_{1-x}$ series hampered reliable KFT results. All samples of this mixing series were dried under identical conditions, i.e., for 24 h in nitrogen atmosphere at 373 K and for 48 h in reduced pressure at 353 K. Subsequently, the samples were sealed into the DSC aluminum pans in dried $N_2$-atmosphere. Comparing the measured $T_g$ with previous results[27] the water content of these samples can be determined to be significant less than 1 mol%. For the dielectric measurements < 3 GHz all samples were dried already in the employed cryostats, due to the small sample size no water content can be reported. The expanded spectra up to 40 GHz for $BPY\ BF_4$ was recorded in ambient atmosphere; thus the sample was dried separately, similarly as for the calorimetry measurements.

**Thermal measurements.** For the DSC measurements a DSC 8500 (Perkin Elmer) was used. The samples were cooled to 100 K and subsequently heated back to room temperature at 10 K/min. Aluminum pans, which were hermetically sealed after filling in the sample, were used for all measurements. The glass-transition temperatures were taken at the onset of the transition steps in the heating traces.

**Dielectric measurements.** The complex permittivity $\varepsilon^* = \varepsilon' - i\varepsilon''$ and the real part of the conductivity $\sigma'$ at frequencies $1\ Hz \leq \nu \leq 1\ MHz$ were determined using a frequency-response analyzer (Novocontrol alpha-Analyzer) and at $1\ MHz \leq \nu \leq 3\ GHz$ employing a coaxial reflection impedance analyzer (Agilent E4991A or Keysight E4991B). All samples were measured in steel-plate capacitors. For sample cooling between 170 and 370 K, a Novocontrol Quatro Cryosystem was utilized. For $BPY\ BF_4$, the spectrum was additionally expanded to 40 GHz using a coaxial reflection measurement with an open-ended sensor (Agilent E8363B PNA Series Network Analyzer with 85070E Dielecric Probe Kit). Sample heating from 300 to 370 K was done by an Eppendorf Thermomixer Comfort with an additional external temperature sensor.

**Acknowledgements**
We thank M. Aumüller and M. Weiss for performing parts of the dielectric measurements and parts of the DSC measurements. This work was supported by the BMBF via ENREKON 03EK3015 and by the Bavarian graduate school "Resource strategy concepts for sustainable energy systems" of the Institute of Materials Resource Management (MRM) of the University of Augsburg.

**Author contributions**
P.S. and S.K. initiated the research. S.K. supervised the project. D.R. and M.W. performed the DSC and E.T. the dielectric measurements. P.S. and S.K. wrote the paper with contributions from E.T. and A.L. All authors discussed the results and commented on the manuscript.

**Additional information**
Competing financial interests: The authors declare no competing financial interests.